\documentstyle[aps,preprint,prl]{revtex}

\begin{document}

\title{Spin-Charge Separated Luttinger Liquid in Arbitrary Spatial Dimensions
}

\author{Krzysztof Byczuk$^{a}$ and Jozef Spa\l ek$^{b,a}$
\footnote{E-mail: byczuk@fuw.edu.pl, ufspalek@if.uj.edu.pl}}

\address{(a) Institute of Theoretical Physics, Warsaw University,
ul. Ho\.za 69, 00-681 Warszawa, Poland \\
(b) Institute of Physics, Jagiellonian University, ul. Reymonta 4, 30-059
Krak\'ow, Poland }

\maketitle

\begin{abstract}
A  model with a singular forward scattering amplitude
for particles with opposite spins in d  spatial dimensions is proposed
and solved by  using the bosonization transformation.
This interacting potential leads to the  spin-charge
separation.
Thermal properties at low temperature for this Luttinger liquid are
discussed.
Also, the explicit form of the single-electron Green function is found; it has
square-root branch cut.
New fermion field operators are defined;
they describe holons and spinons as  the elementary excitations.
Their single particle Green
functions possess pseudoparticle properties.
Using these operators the spin-charge separated Hamiltonian for
an ideal gases of holons and spinons is derived
and reflects an inverse (fermionization) transformation.

\end{abstract}


It was suggested  \cite{and1} that the properties of normal state of
high-temperature superconductors are properly described
by Luttinger liquid,  where the spin and charge degrees of
freedom are separated.
In one-dimensional systems this phenomenon is well understood \cite{rew}.
However, in two and three dimensions the present understanding
of the spin-charge
separation is rather poor.
In this letter we formulate and solve exactly a  d-dimensional model
exhibiting  the spin-charge separation, as well as
discuss its thermal and dynamic
properties.

A natural approach to study spin-charge decoupling phenomena
is the bosonization transformation, generalized recently
to the multidimensional space situation \cite{hal}.
Here we adopt the operator version of the bosonization developed
in Ref.\cite{how}.
The starting assumption in this method is the existence of the Fermi
surface (FS) defined as a collection of points at which the momentum
distribution function has singularities at zero temperature ($T=0$).
These points are parameterized by vectors $\bf S$ and ${\bf T}$, which label
a finite and a locally flat (rectangular in shape) mesh of grid points on
FS with spacing  $\Lambda \ll k_F$ between them \cite{how,hal}.
Introducing  coarse-grained
density fluctuation operators $J_{\sigma}({\bf S},{\bf q})$, defined in boxes
centered at each FS point and having surface area $\Lambda^{d-1}$
and the thicknesses $\lambda/2$
both above and below it, one can transform the effective Hamiltonian for
interacting fermions into an effective  Hamiltonian for free bosons.
Explicitly, it takes the general form
\begin{equation}
H = \frac{1}{2} \sum_{{\bf S}, {\bf T}} \sum_{\bf q} \sum_{\sigma \sigma'}
\Gamma_{\sigma \sigma'} ({\bf S}, {\bf T}, {\bf q}) J_{\sigma}({\bf S}, {\bf
q})J_{\sigma'}({\bf T}, - {\bf q}),
\label{e7}
\end{equation}
where
$
\Gamma_{\sigma \sigma'} ({\bf S}, {\bf T}, {\bf q}) = v_F( {\bf S})
\frac{1}{\Omega} \delta_{\sigma, \sigma'}
\delta^{d-1}_{{\bf S}, {\bf T}} +
\frac{1}{L^d} V_{\sigma \sigma'}({\bf S}, {\bf T}, {\bf q})
$
is the positive defined matrix element.
The first term  corresponds to the kinetic energy part of the original
fermionic Hamiltonian with linearized dispersion relation close to  FS,
whereas the second term is the effective (low-energy) interaction between
the particles  with spins $\sigma$ and $\sigma'$.
The geometrical factor $\Omega = \Lambda^{d-1} (\frac{L}{2 \pi})^d$ depends
on the system dimension $d$.

The explicit expression for $V_{\sigma \sigma'}({\bf S},{\bf T}, {\bf q})$ is
generally
  derived
by transforming out the high energy
modes in the fermionic Hamiltonian \cite{shankar}.
Obviously, this procedure  can also change the Fermi velocity $v_F({\bf S})$.
Therefore, we take $v_F({\bf S})$ as an effective value  obtained after
removing
the high-energy degrees of freedom.
As shown below,
we can characterize the universal properties of fermions  knowing only
the asymptotic behavior of the interaction potential
 $V_{\sigma \sigma'}({\bf S}, {\bf T}, {\bf q})$ in the thermodynamic limit.

If the system is invariant under the time reversal, the interaction part
must be explicitly symmetric under this operation, which means
that
$
V_{\sigma \sigma'}({\bf S}, {\bf T}, {\bf q}) = V_{\bar{\sigma}
\bar{\sigma}'}(-{\bf S}, -{\bf T}, - {\bf q})
$.
Furthermore, if  FS is also invariant under the reflections
${\bf S} \rightarrow - {\bf S}$ etc., the last condition becomes
$
V_{\sigma \sigma'}({\bf S}, {\bf T}, {\bf q}) = V_{\bar{\sigma}
\bar{\sigma}'}({\bf S}, {\bf T},  {\bf q})
$.
In that case $V_{\sigma \sigma'}({\bf S}, {\bf T}, {\bf q})$ depends only on
the relative
orientation of the spins $\sigma$ and $\sigma'$; there are only two independent
components:  $V_{\sigma \sigma}$ for parallel spins and
 $V_{\sigma \bar{\sigma}}$ for antiparallel spins.
It is convenient to introduce the symmetric and the antisymmetric
combinations:
$
V^{c,s}({\bf S},{\bf T},{\bf q}) \equiv \frac{1}{2} (V_{\sigma \sigma}({\bf S},
{\bf T}, {\bf q})
\pm V_{\sigma \bar{\sigma}}({\bf S}, {\bf T}, {\bf q}) ),
$
where $c$ and $s$ superscripts corresponds to "$\pm$" signs, respectively.
Correspondingly, we define the currents
$
J_{c,s}
({\bf S}, {\bf q}) \equiv \frac{1}{\sqrt{2}} ( J_{\uparrow}({\bf S}, {\bf q})
\pm J_{\downarrow}({\bf S}, {\bf q})),
$
which describe the charge and the spin density fluctuations, respectively.
Then, the original Hamiltonian (\ref{e7}) takes the form
\begin{equation}
H = \sum_{\alpha = c,s}
\frac{1}{2} \sum_{{\bf S}} v_F({\bf S}) \frac{1}{\Omega} \sum_{{\bf q}}
J_{\alpha}({\bf S}, {\bf q}) J_{\alpha}({\bf S}, - {\bf q}) +
 \frac{1}{L^d} \sum_{{\bf S}, {\bf T}} \sum_{{\bf q}}
V^{\alpha}({\bf S}, {\bf T}, {\bf q}) J_{\alpha}({\bf S}, {\bf q}) J_{\alpha}(
{\bf T}, -{\bf q}).
\label{e15}
\end{equation}
The $\alpha =c$ term
describes the dynamics of the charge density fluctuations in the system,
whereas the $\alpha = s$ term deals with the longitudinal
spin density fluctuations.
There are no terms which mix the  degrees of freedom
(i.e. $ \sim J_c \cdot J_s $) because the Hamiltonian is assumed to be
 invariant
under the spin flip (i.e. $J_c \rightarrow J_c$, $J_s \rightarrow - J_s$).
We can check out that for the noninteracting case, i.e. for $V^{\alpha}\equiv
0$,
both the spin and the charge density fluctuations propagate
 with the same velocity
$v_F({\bf S})$.
The commutation relation for density fluctuation
operators  take
the following form
$
\left[
J_{\alpha}({\bf S}, {\bf q}) ,  J_{\beta}({\bf T}, {\bf p})
\right] = \delta_{\alpha \beta} \delta^{d-1}_{{\bf S}, {\bf T}} \delta^d_{ {\bf
p} + {\bf q},0}\:
\Omega \: {\bf q} \cdot \hat{n}_{\bf S},
$
where $\alpha , \beta = c,s$.
Thus,
the two branches of density fluctuations are independent of each other.
The commutation relations become equivalent to those obeyed by the bosonic
harmonic-oscillator creation and annihilation operators after rescaling them
by the  factor on the right hand side, i.e. by defining
the creation ($a_{\alpha}^+$) and the annihilation ($a_{\alpha}$)
operators according to
\begin{equation}
J_{\alpha}({\bf S}, {\bf q}) = \theta( \hat{n}_{{\bf S}} \cdot {\bf q})
\sqrt{\Omega \hat{n}_{{\bf S}} \cdot {\bf q}}
\; a_{\alpha}({\bf S}, {\bf q}) +
\theta(- \hat{n} _{\bf S} \cdot {\bf q}) \sqrt{-\Omega \hat{n}_{{\bf S}} \cdot
{\bf q}}
\; a_{\alpha}^+({\bf S},- {\bf q}),
\end{equation}
where  $\theta(x) $ is the step function.
Then
$
\left[
a_{\alpha}({\bf S}, {\bf q}), a_{\beta}^+({\bf T}, {\bf p})
\right]
= \delta^{d-1}_{{\bf S}, {\bf T}} \delta_{{\bf p},{\bf q}}^d \delta_{\alpha
\beta}.
$
In  terms of the bosonic harmonic-oscillator creation and annihilation
operators the Hamiltonian (\ref{e15}) is
\begin{equation}
H = \sum_{\alpha = c,s} \sum_{{\bf S}, {\bf T}} \frac{1}{2}
\sum_{{\bf q}} \left\{ \theta( \hat{n}_{{\bf S}} \cdot {\bf q})
\theta(\hat{n}_{{\bf T}}\cdot {\bf q})
\sqrt{ ({\bf v}_F({\bf S})  \cdot {\bf q}) ({\bf v}_F({\bf T}) \cdot {\bf q})}
\; \times
\right.
\label{e22}
\end{equation}
$$
\left( \delta_{{\bf S}, {\bf T}}^{d-1} +
 \frac{\Lambda^{d-1}}{(2 \pi)^d}
\frac{V^{\alpha}({\bf S},{\bf T},{\bf q})}{\sqrt{v_F({\bf S})v_F({\bf T})}}
\right)
a_{\alpha}^+({\bf S},{\bf q}) \: a_{\alpha}({\bf T}, {\bf q}) \; +
$$
$$
 \theta(- \hat{n}_{{\bf S}} \cdot {\bf q}) \theta(-\hat{n}_{{\bf T}}\cdot {\bf
q})
\sqrt{ (-{\bf v}_F({\bf S})  \cdot {\bf q}) (-{\bf v}_F({\bf T}) \cdot {\bf
q})} \; \times
$$
$$
\left( \delta_{{\bf S}, {\bf T}}^{d-1}  +
 \frac{\Lambda^{d-1} }{(2 \pi)^d}
\frac{V^{\alpha}({\bf S},{\bf T},{\bf q})}{\sqrt{v_F({\bf S})v_F({\bf T})}}
\right)
a_{\alpha}^+({\bf S},-{\bf q})\: a_{\alpha}({\bf T},- {\bf q}) \; +
$$
$$
 \theta( \hat{n}_{{\bf S}} \cdot {\bf q}) \theta(-\hat{n}_{{\bf T}}\cdot {\bf
q})
\sqrt{ ({\bf v}_F({\bf S})  \cdot {\bf q}) (-{\bf v}_F({\bf T}) \cdot {\bf q})}
\left(   \frac{\Lambda^{d-1}}{(2 \pi)^d}
\frac{V^{\alpha}({\bf S},{\bf T},{\bf q})}{\sqrt{v_F({\bf S})v_F({\bf T})}}
\right)
a_{\alpha}({\bf S},{\bf q})\: a_{\alpha}({\bf T}, -{\bf q}) \; +
$$
$$
\theta(- \hat{n}_{{\bf S}} \cdot {\bf q}) \theta(\hat{n}_{{\bf T}}\cdot {\bf
q})
\sqrt{ (-{\bf v}_F({\bf S})  \cdot {\bf q}) ({\bf v}_F({\bf T}) \cdot {\bf q})}
\left.
\left(  \frac{ \Lambda^{d-1}}{(2 \pi)^d}
\frac{V^{\alpha}({\bf S},{\bf T},{\bf q})}{\sqrt{v_F({\bf S})v_F({\bf T})}}
\right)
a_{\alpha}^+({\bf S},-{\bf q})\: a_{\alpha}^+({\bf T}, {\bf q})
\right\}.
$$

We have shown previously \cite{my,phd}
that a universal behavior of the system takes place and depends
on how the interaction part of the effective Hamiltonian  (\ref{e22})
behaves in the scaling (thermodynamic) limit $\Lambda \rightarrow 0$.
If the interaction  in this limit has a singular power-law
 behavior, i.e. $V_{\sigma,\bar{\sigma}}({\bf S},{\bf T},{\bf q})$
 scales for ${\bf S} \rightarrow {\bf T}$ as
$\Lambda^{d-1} / \Lambda^{\eta}$, then: (i) for $\eta < d-1$ the Landau
Fermi liquid (FL) fixed point is stable; (ii) for $\eta > d-1$ the statistical
spin liquid (SSL)
is the stable fixed point; (iii) the case  $\eta = d-1$ leads to the
Luttinger liquid (LL) type of behavior.
Properties of FL in $d=3$ and $2$ are well known \cite{lfl}.
SSL was discussed in Refs.\cite{ssl,my,phd}.
Here we concentrate on  the properties of the spin-charge separated LL
in an arbitrary spatial dimension.

To examine the basic properties of Luttinger liquid we assume
the following form of the effective interaction in (\ref{e7}):
\begin{equation}
V_{\uparrow \uparrow} ({\bf S}, {\bf T}, {\bf q}) = f_{\uparrow \uparrow}({\bf
S}, {\bf T}, {\bf q}),
\end{equation}
and
\begin{equation}
V_{\uparrow \downarrow} = \left\{
\begin{array}{ccc}
\frac{1}{\Lambda^{d-1}} \tilde{g}({\bf S}, {\bf q}) & for & {\bf S} = {\bf T}
\\
f_{\uparrow \downarrow} ({\bf S}, {\bf T}, {\bf q}) & for & {\bf S} \neq {\bf
T} ,
\end{array}
\right.
\label{e34}
\end{equation}
where $f_{\sigma \sigma'} ({\bf S}, {\bf T}, {\bf q})$ and
$\tilde{g}({\bf S}, {\bf q}) $
are nonsingular functions.
In this manner,  we model the situation with a divergent
forward scattering amplitude
(taking place  for ${\bf S} = {\bf T}$ and ${\bf q}=0$), which
leads to the LL fixed point.
Also, since we examine only the low-energy limit, we can expand the nonsingular
part according to:
$
\tilde{g}({\bf S}, {\bf q}) = g_0({\bf S}) + g({\bf S}) \hat{n}_{{\bf S}} \cdot
{\bf q} + ...
$, and omit the higher-order terms.
Substituting this expansion into (\ref{e34}) we obtain
 in the thermodynamic limit
($\Lambda \rightarrow 0$) the  Hamiltonian
with two branches of free bosons excitations, each
with a different form of the kinetic
energy.
Namely, defining
$
v_F^{c,s} ({\bf S}) \equiv v_F({\bf S}) \pm g({\bf S}),
$
the Hamiltonian (\ref{e22}) simplifies to
\begin{equation}
H = \sum_{\alpha = c , s}
\sum_{{\bf S}, {\bf q}> 0} ({\bf v}_F^{\alpha}({\bf S}) \cdot {\bf q} ) \;
a_{\alpha}^+ ({\bf S}, {\bf q})
\; a_{\alpha}({\bf S}, {\bf q}).
\label{e40}
\end{equation}
The charge and the spin fluctuations propagate in the system with
different velocities and express the separation of the corresponding
degrees of freedom.
One of the velocities diminishes and the other increases.
In the extreme case one of them vanishes transforming into a soft mode,
signalling a phase transition  in the corresponding
channel, $c$ or $s$.
The character of this transition will be determined by the regular part.
The spin-charge decoupling takes always place  in the
one-dimensional systems of
 interacting fermions in the low energy limit \cite{rew}.
In a system of  higher dimension the potential must be singular.
Obviously,  this is a simplified model.
More appealing form would be
$
V_{\uparrow \downarrow}({\bf S}, {\bf T}, {\bf q}) \sim \frac{1}{
|{\bf S} - {\bf T}|^{\eta} + |{\bf q}|^{\eta} },
$
which clearly diverges in  the forward direction (for ${\bf q} \rightarrow 0$).
Such a singular effective potential in $d=2$ was discussed in
Refs.\cite{and2,fqhe,stamp} in the context of the high-temperature
superconductors and the fractional quantum Hall effect.
Our choice (\ref{e34}) is modeled
by  the most divergent term of this potential,
and physically means that the fermions  with antiparalel spins interact
through the forward scattering processes along the radial FS direction only.

Since the Hamiltonian (\ref{e40})
is diagonal we can calculate the internal energy
of the system and then the specific heat.
The energy is
\begin{equation}
U = E_0 + \sum_{\alpha=c,s}
\sum_{{\bf S}} \sum_{{\bf q}} \frac{ {\bf v}_F^{\alpha}({\bf S}) \cdot {\bf q}}
{e^{\beta {\bf v}_F^{\alpha}({\bf S}) \cdot {\bf q}} - 1},
\label{e42}
\end{equation}
where we utilize the fact that now we are dealing with bosons. $E_0$ is the
ground state energy of the initial noninteracting system.
This term must be incorporated because in the bosonization procedure
 the energy  is measured with respect to the Fermi energy.
Also, the
 chemical potential $\mu$ does not appear in (\ref{e42}) because the number
of bosons is not conserved.
In other words, those fields describe the system  particle-hole
excitations.
This is also the reason why those bosons cannot condense.
For an isotropic system the velocities do not depend on the index labeling the
point of  FS, i.e.
 $v_F^{\alpha}({\bf S}) = v_F^{\alpha}$, and we find that
$
U = \frac{\pi^2}{6} (k_B T)^2 \left( \frac{L}{2 \pi} \right) ^d
\frac{d \pi^{d/2}}{\Gamma(d/2 + 1)} k_F^{d-1} \sum_{\alpha}
\frac{1}{v_F^{\alpha}}
+ E_0.
$
Hence, the specific heat is
$
C_V = \frac{1}{L^d} \frac{\partial U}{\partial T} =
\frac{\pi^2}{3} (k_B )^2 T \left( \frac{1}{2 \pi} \right) ^d
\frac{d \pi^{d/2}}{\Gamma(d/2 + 1)} k_F^{d-1} \sum_{\alpha}
\frac{1}{v_F^{\alpha}}.
$
Introducing the density of states for the charge and the spin excitations on
 FS:
$
\rho_{\alpha}(\epsilon_F) = \left( \frac{1}{2 \pi} \right) ^d
\frac{d \pi^{d/2}}{\Gamma(d/2 + 1)} k_F^{d-1}  \frac{1}{v_F^{\alpha}},
$
we have that
$
C_V = \frac{\pi^2}{3} (k_B )^2 T \sum_{\alpha} \rho_{\alpha} (\epsilon_F)
\equiv
\gamma_{LL} T.
$
The spin-charge separated liquid has a linear
specific heat, with
$
\gamma_{LL} \sim \frac{1}{v_F^c} + \frac{1}{v_F^s}.
$
The linear specific heat is thus a general characteristic following
from the existence of the FS,
 independently of the statistical properties  of the
particles.
In the limiting case $v_F^c = v_F^s = v_F$ we recover the FL result.
Similarly, we find the free energy in the form
$
F= E_0 - \frac{\pi^2}{6} (k_BT)^2 \sum_{\alpha} \rho_{\alpha}(\epsilon_F),
$
and the low-$T$ entropy
$
S = \frac{U-F}{T} = \frac{\pi^2}{3} (k_B)^2 T \sum_{\alpha}
\rho_{\alpha}(\epsilon_F),
$
which coincides with the specific heat.
Thus the low-T thermal properties of the
present Luttinger spin-charge separated liquid are very similar to those of
free fermions.
The only difference is in the form of the density of states at
FS.
However, the dynamic properties in the LL case are quite unique, as
we discuss next.

We define  the fermion correlation function as
\begin{equation}
G^>_{\sigma} ({\bf S}, {\bf x}, t > 0) = < \psi_{\sigma}({\bf S}, {\bf x}, t)
\psi_{\sigma}^+ ({\bf S}, 0, 0)>.
\label{e50}
\end{equation}
To derive an explicit form of
this function we substitute the bosonized Fermi field
operators \cite{how}
$
\psi_{\sigma} ({\bf S}, {\bf x}) = \sqrt{\frac{\Omega}{a}} e^{i {\bf k}_{{\bf
S}} \cdot {\bf x}}
e^{i \frac{\sqrt{4 \pi}}{\Omega} \phi_{\sigma} ({\bf S},{\bf x})} \hat{O}({\bf
S}),
$
and utilize the identity
$
e^A e^B = :e^{A+B}: e^{<AB - \frac{1}{2}(A^2 + B^2)>},
$
where $::$ means that $:e^{A+B}:$ is a normal ordered product of operators.
Then, we  have that
$
G^>_{\sigma} ({\bf S}, {\bf x}, t) = \frac{\Omega}{a} exp \left(
{\frac{4 \pi}{\Omega ^2} \frac{1}{2}
G_B^{\sigma}({\bf S}, {\bf x}, t)} \right),
$
where $G_B^{\sigma}$ is expressed via the Bose fields, namely
$
G_B^{\sigma}({\bf S}, {\bf x}, t) = < (\phi_c({\bf S}, {\bf x}, t) + \sigma
\phi_s({\bf S}, {\bf x}, t))
(\phi_c({\bf S}, 0, 0) + \sigma \phi_s({\bf S}, 0, 0))> -
<(\phi_c({\bf S}, 0, 0) + \sigma \phi_s({\bf S}, 0, 0))^2 >.
$
Next, using the Heisenberg representation for the boson field operators
$\phi_{\alpha} ({\bf S},{\bf x},t)$, and subsequently
decomposing them into the Fourier components, we find the explicit form of
the boson correlation function:
$
G_B^{\sigma}({\bf S}, {\bf x}, t) = - \frac{\Omega^2}{4 \pi}
\ln \left( \frac{ (\hat{n}_{{\bf S}}\cdot {\bf x} - v_F^c t+ ia) (
\hat{n}_{{\bf S}} \cdot {\bf x} -
v_F^s t + ia)}{ (ia)^2} \right) .
$
Hence the fermion correlation function is
\begin{equation}
G_{\sigma}^> ({\bf S}, {\bf x}, t) = i \Omega \frac{ e^{i {\bf k}_{{\bf S}}
\cdot {\bf x}}
}{ \sqrt{\hat{n}_{{\bf S}}\cdot {\bf x} - v_F^c t + ia }
\sqrt{ \hat{n}_{{\bf S}} \cdot {\bf x} -
v_F^s t + ia }
}.
\label{e55}
\end{equation}
It is independent of the spin index $\sigma$.
We see that instead of a quasiparticle pole, taking place  in the FL case,
  we have now a branch cut ranging from
$v_F^s$ to $v_F^c$.
This branch cut survives when we transform $G^>_{\sigma}({\bf S}, {\bf x}, t)$
into
$G^>_{\sigma}({\bf S}, {\bf k}, \omega)$
as can be easily seen by decomposing the argument into normal and
transverse parts: ${\bf k} \cdot {\bf x} =
k ( \hat{n}_{{\bf S}} \cdot {\bf x} + {\bf t} \cdot {\bf x})$,
where ${\bf t} k$ is the transverse
part of ${\bf k}$, and noting  that the part
$\hat{n}_{{\bf S}} \cdot {\bf x}$ can be transformed
in the same manner, as in the $d=1$ case \cite{and3}.
The
${\bf t} \cdot {\bf x}$ part has a trivial form.
This means that the analytic character of the LL Green function
is the same in both $d=1$ and $d>1$ cases.
This universal character follows from the relation
$
G^>({\bf S}, {\bf k}, \omega) \equiv G^> ( {\bf S}, \hat{n}_{{\bf S}}\cdot {\bf
k}, \omega)
\delta_{
{\bf t} \cdot {\bf k}, 0}$.
This very significant result  tells us also that there are no
quasiparticle excitations having a direct relation to the noninteracting
particles in this charge-spin separated system.
In other words, when we put a single electron forming a wave packet
on  FS, it dissociates into many wave packets propagating
with velocities ranging form $v_F^s$ to $v_F^c$.
This nonperturbative result means that there is no one-to-one correspondence
between the dynamics of LL liquid system
and the system of non-interacting fermions.
Note that the distribution function $\bar{n}_{\bf k}$ is in our model
situation at $T=0$ a step function with a jump at $k_F$, as in the FL case.
The inclusion the
nonsingular part of the interaction will change this step distribution for
a finite-volume system.

The fundamental question arises if we can still define a proper
fermionic pseudoparticles in this spin-charge separated liquid.
To construct such a state we consider
 the field operators $\psi_c$ and $\psi_s$, defined through
the following fermionization transformation
\begin{equation}
\psi_{c,s} ({\bf S}, {\bf x}) = \sqrt{\frac{\Omega}{a}} e^{i {\bf k}_{{\bf S}}
\cdot {\bf x}}
e^{i \frac{\sqrt{4 \pi}}{\Omega} \phi_{c,s} ({\bf S},{\bf x})} \hat{O}({\bf
S}).
\label{e58}
\end{equation}
The operators $\psi_{\alpha}({\bf S},{\bf x})$ obey proper anticommutation
relations.
By the procedure similar to that employed in deriving (\ref{e55}),
we now have  the correlation function in the
new fermionic variables in the form
\begin{equation}
G^>_{\alpha} ( {\bf S}, {\bf x}, t>0)
\equiv <\psi_{\alpha}({\bf S}, {\bf x}, t) \psi_{\alpha}^+({\bf S},0,0)>
= i \Omega \frac{ e^{ i {\bf k}_{{\bf S}} \cdot {\bf x} }
}{ \hat{n}_{{\bf S}} \cdot {\bf x} - v_F^{\alpha} t + ia
},
\end{equation}
which has  simple poles.
Also, one can show that
$[ \psi^+_{\alpha} ({\bf S},{\bf x}), \int d {\bf y} J_{\alpha}({\bf S}, {\bf
y}) ]
=  \psi^+_{\alpha}({\bf S}, {\bf x})$,
which means that $\psi^+_{\alpha}( {\bf S},{\bf x})$
 changes the total number of either   charge
$(\alpha=c)$ or   spin $(\alpha=s)$ of the system
at the FS point ${\bf S}$ by one unit.
Hence, the operators (\ref{e58})
represent new
fermionic pseudoparticles
for the interacting non-Fermi liquid.
Those single-particle excitations are called the {\bf holons} for $\alpha = c$
and the {\bf spinons} for $\alpha =s$.
They are the only single-particle excitations across  FS
in our model system.
Since they obey the fermion statistics, the contribution to the specific heat
is of the fermionic type.
In other words, the spin-charge  separated liquid is composed of
 ideal gases of spinons
and holons.
They represent  the exact eigenstates of the system.
Effectively, we have the following Hamiltonian for noninteracting
fermion pseudoparticles
\begin{equation}
H = \sum_{\alpha = c,s} \sum_{{\bf S}} v_F^{\alpha}
({\bf S}) \int d {\bf x} \; \psi^+_{\alpha}({\bf S}, {\bf x}) \left(
\frac{\hat{n}_{{\bf S}} \cdot \nabla}{i} \right) \psi_{\alpha}({\bf S}, {\bf
x}) .
\label{nonham}
\end{equation}
Since we have treated the interaction  between fermions exactly
in the thermodynamic limit ($\Lambda \rightarrow 0$),
the spectrum of our new noninteracting Hamiltonian (\ref{nonham}) is
exactly
the same as that of the interacting Hamiltonian (\ref{e7}).
To conclude, the two branches of elementary excitations in this LL
 can be represented either as boson (collective) or as fermion
(single-particle) excitations.

Having determined  the excitation spectrum of the singular part
of interaction, which resulted in the spin-charge separated liquid,
we can now include in (\ref{nonham})
the nonsingular parts  of the interaction,
transformed to the fermionic  representation (\ref{e58}).
Since for the former part we determined an exact state of the system,
 we can now treat the residual interaction among them as a perturbation,
i.e. regard the system of interacting holons and spinons as being in one-to-one
correspondence to the system of noninteracting holons and spinons.
This statement can be proved by referring to
 the Gell-Mann and Low theorem which means that the evolution operator is
well defined in all orders,
 since the
interactions among the holons and the spinons are nonsingular functions.
In other words, the adiabatic "switching on" procedure is justified.
Here  we discuss only a  semiclassical approach to the interacting
holons and spinons in the bosonic language.
The equations of motion for  $J_{\alpha}({\bf S},{\bf q})$ operators are
\begin{equation}
i \frac{\partial}{\partial t} J_{\alpha}({\bf S}, {\bf q},t) =
{\bf v}_F^{\alpha}({\bf S}) \cdot {\bf q} \; J_{\alpha}({\bf S},{\bf q},t) +
{\bf q} \cdot \hat{n}_{{\bf S}} \; \Lambda^{d-1} \left( \frac{1}{2\pi}
\right)^d \sum_{{\bf T}} f_{\alpha}({\bf S} - {\bf T}) J_{\alpha}({\bf S}, {\bf
q},t),
\end{equation}
where we supposed that the interaction $f_{\alpha}$ does not depend on ${\bf
q}$.
In the semiclassical approach we take the expectation value of $J_{\alpha}$,
i.e.
define
$
u_{\alpha}({\bf S},{\bf q},t) \equiv < J_{\alpha}({\bf S},{\bf q},t)>$.
This quantity measures the shape deformation of FS
with respect to the ground state form.
Furthermore, focusing our attention on the single Fourier mode
$u_{\alpha}(t)  = e^{-i \omega t}u_{\alpha}$, we find the integral equation for
the
collective excitation spectrum amplitude
$u_{\alpha} \equiv u_{\alpha} (\tilde{\Omega},{\bf q})$, in the form
\begin{equation}
(v_F^{\alpha} q \cos \theta - \omega ) u_{\alpha} (\tilde{\Omega}, q) =
q v_F^{\alpha} \cos \theta \int \frac{d \tilde{\Omega}'}{S_d}
F_{\alpha}(\tilde{\Omega}, \tilde{\Omega}')
u_{\alpha}(\tilde{\Omega}', q),
\label{e63}
\end{equation}
where $S_d = \int d \tilde{\Omega}$, $F_{\alpha}(\tilde{\Omega},
\tilde{ \Omega}') = \rho(\epsilon_F)
f_{\alpha}(\tilde{\Omega},
\tilde{\Omega}')$, and $\tilde{\Omega} $ is the solid angle.
This is an
equation of motion for either the holon ($\alpha =c$) or the
 spinon ($\alpha=s$) sound-wave amplitudes.
The introduced bosons represent the sound waves propagating around
FS, here characterized  by $\omega$ and ${\bf q}$.
Eq.(\ref{e63}) gives both the stable solution for collective modes and
the solution with the imaginary frequency.
Solution of Eq.(\ref{e63}) is analogical to that
considered in the FL theory
\cite{lfl};
it will not be discussed in detail here.

In summary, we presented a model  with
the
spin-charge separation in the space of  arbitrary  dimensions,
as well as have  discussed
some of its basic properties.
The next step would require a careful analysis of the nonsingular part
of the interaction for finite-volume systems.
In particular, the most important question is to construct a
 theory of interacting holons and spinons at low energies (in the spirit
of the Landau FL), including also the effects coming from the
presence of an applied magnetic field.
One should  also examine the stability of this liquid against
charge or spin-density
wave formation.

The paper was supported by the Committee of Scientific Research (KBN) of
Poland.
The work was performed in part at Purdue University (U.S.A.), where it was
supported by the MISCON Grant No. DE-FG 02-90 ER 45427, and by the NSF Grant
No. INT. 93-08323.

\end{document}